\documentclass[aps,preprint,nofootinbib,superscriptaddress]{revtex4}

\usepackage{amssymb}

\usepackage{epsfig}
\usepackage[bookmarksnumbered,bookmarksopen,colorlinks,citecolor=blue,linkcolor=blue]{hyperref}

\begin{document}

\title{ Fission barriers with Weizs\"acker-Skyrme mass model }

\author{Ning Wang}
\email{wangning@gxnu.edu.cn}\affiliation{ Department of Physics,
Guangxi Normal University, Guilin 541004, People's Republic of
China }
\affiliation{ Guangxi Key Laboratory of Nuclear Physics and Technology, Guilin 541004, People's Republic of
China }

\author{Min Liu}
\email{liumin@gxnu.edu.cn}\affiliation{ Department of Physics,
Guangxi Normal University, Guilin 541004, People's Republic of
China }
\affiliation{ Guangxi Key Laboratory of Nuclear Physics and Technology, Guilin 541004, People's Republic of
China }

\begin{abstract}
Based on Weizs\"acker-Skyrme (WS4) mass model, the fission barriers of nuclei are systematically studied. Considering the shell corrections, the macroscopic deformation energy and a phenomenological residual correction, the fission barrier heights for nuclei with $Z\geqslant 82$ can be well described, with an rms deviation of 0.481 MeV with respect to 71 empirical barrier heights. In addition to the shell correction at the ground state, the shell correction at the saddle point and its relative value are also important for both deformed nuclei and spherical nuclei. The fission barriers for nuclei far from the $\beta$-stability line and super-heavy nuclei are also predicted with the proposed approach.

\end{abstract}
\maketitle

\begin{center}
\textbf{I. INTRODUCTION}
\end{center}

Studies of nuclear fission are of great interest \cite{Co63,Bj80,Cap09,And18,Liu96,Deni22}. As one of the key and sensitive physical inputs, fission barriers of nuclei are frequently used in the study of nuclear physics \cite{Fan00,Zhao15,Zhang19,Zhang21,Zhou21,Hof16}, reactor physics \cite{Wag77} and nuclear astrophysics \cite{Pan05,Gor15}. In the synthesis of super-heavy nuclei (SHN) through fusion reactions, the prediction of the evaporation-residue cross section for SHN is strongly dependent on the fission barrier height of the compound nuclei adopted in the calculations \cite{Wang11,Adam20,Zhang23,Nov20,Lv21}. An 1-MeV shift of the barrier height may result in a change of the calculated cross section of 3n or 4n reaction by $2-3$ orders of magnitude \cite{Siwek12,Nas11}. Considering the complex of the fission process where not only large-scale collective participation of the nucleons is witnessed but also superdeformed shapes of nuclei are encountered, accurate description of the fission barrier is of great interest and challenging.

The fission barriers of nuclei can be described with some nuclear mass models, such as the macroscopic-microscopic models \cite{Pomo03,Kow10,Moll15} and the microscopic models based on the Skyrme energy density functionals  \cite{Mam98,Sam05,Pei16} or the covariant density functionals \cite{Abu12,Lu14,Zhou16}, in which the model parameters are usually determined by the nuclear properties at the ground state. For unmeasured SHN, the uncertainty of the predicted barrier heights from these different models could reach a few MeV \cite{Nas11,Kow10}. Considering the strong influence of the barrier height on the prediction of cross section as mentioned above, it is therefore necessary to improve the model accuracy for describing the fission barriers of unstable nuclei.

Inspired by the Skyrme energy-density functional, the Weizs\"acker-Skyrme (WS4) mass model was proposed \cite{WS4}, with which the known masses can be reproduced with an rms error of $\sim0.3$ MeV \cite{Zhao22} and the known $\alpha$-decay energies of SHN can be reproduced with deviations smaller than 0.5 MeV \cite{Ogan15,Wang15,Pei24}. It would be therefore interesting to apply the WS4 model for describing the fission barrier. In this work, we attempt to systematically calculate the fission barrier height based on the WS4 model, considering the shell correction at the ground state and that at the saddle point.

\begin{center}
\textbf{ II. THEORETICAL FRAMEWORKS  }
\end{center}

In the macroscopic-microscopic model, the fission barrier height of a nucleus at zero temperature is expressed as the difference between the energy of the nucleus at saddle point $E_{\rm sad}$ and that at its ground state $E_{\rm gs}$,
\begin{eqnarray}
B_{\rm f}=E_{\rm sad}-E_{\rm gs} = (E_{\rm sad}^{\rm mac}-E_{\rm gs}^{\rm mac})+(U_{\rm sad}-U_{\rm gs})+\Delta B.
\end{eqnarray}
Here, $E_{\rm sad}^{\rm mac}$ and $E_{\rm gs}^{\rm mac}$ denote the macroscopic energy at the saddle point and that at the ground state, respectively. $U_{\rm sad}$ and $U_{\rm gs}$ denote the corresponding shell corrections. $\Delta B$ denotes the residual correction. With the macroscopic fission barrier height $B_{\rm f}^{\rm mac}=E_{\rm sad}^{\rm mac}-E_0^{\rm mac}$ and the macroscopic deformation energy $B_{\rm def}=E_{\rm gs}^{\rm mac}-E_0^{\rm mac}$, the fission barrier height can be re-written as
\begin{eqnarray}
B_{\rm f}= B_{\rm f}^{\rm mac}-U_{\rm gs}+(U_{\rm sad}-B_{\rm def})+\Delta B.
\end{eqnarray}
For spherical nuclei, the barrier height can be estimated with $B_{\rm f}^{\rm (0)} = B_{\rm f}^{\rm mac}-U_{\rm gs}$ if neglecting the saddle point shell correction and the residual correction.

\begin{figure}
\includegraphics[angle=0,width=1\textwidth]{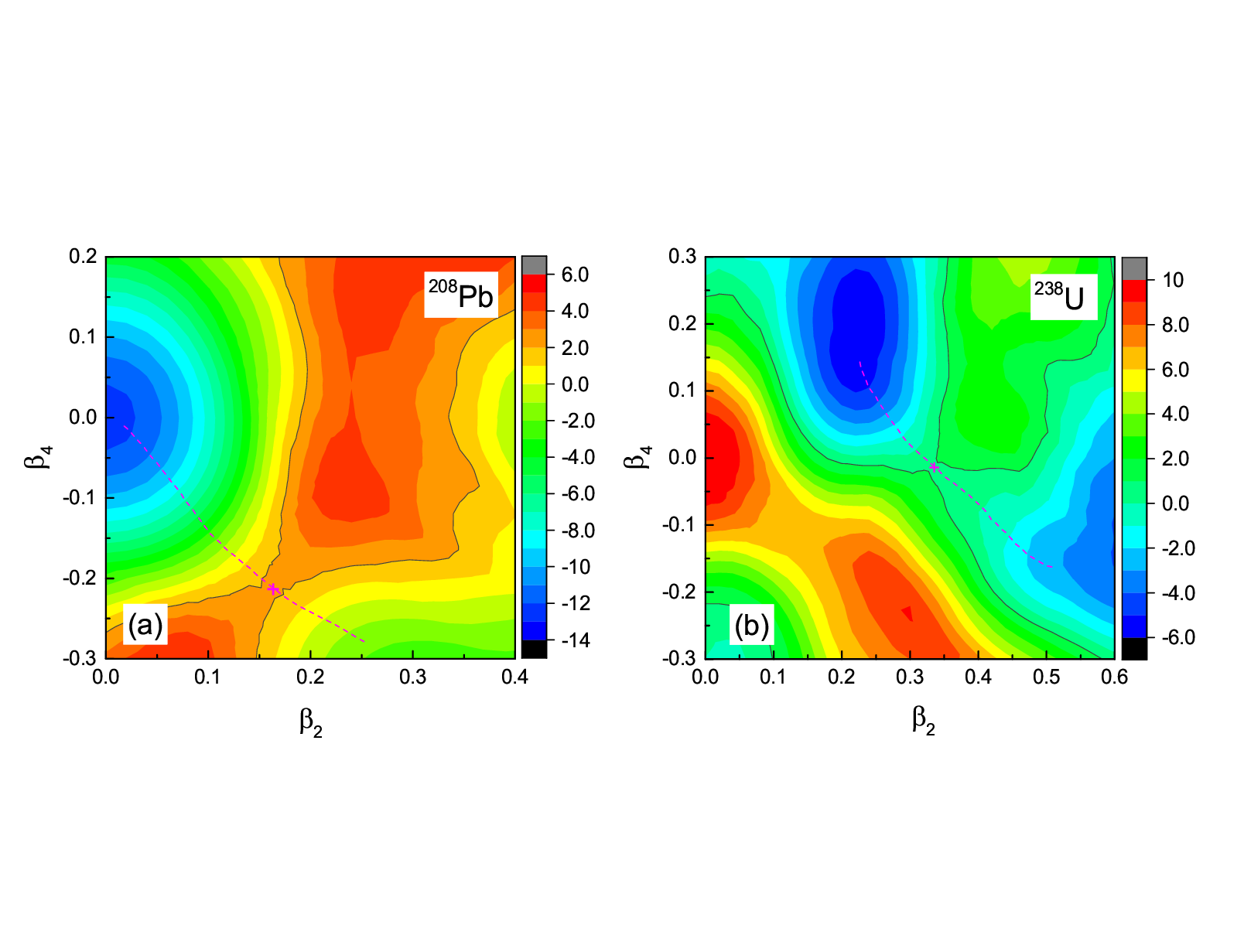}
\caption{(Color online) Contour plot of the shell corrections for $^{208}$Pb and $^{238}$U. The dashed curves denote the possible fission paths. The crosses denote the positions of the saddle points.   }
\end{figure}

Following Cohen-Swiatecki's formula \cite{Co63}, the macroscopic fission barrier height is expressed as,
\begin{eqnarray}
B_{\rm f}^{\rm mac}= \left\{
\begin{array} {r@{\quad:\quad}l}
  0.38(3/4-x)E_s  &   1/3 < x \leqslant 2/3    \\
  0.88(1-x)^3E_s  &   2/3 < x \leqslant 1    \\
 \end{array} \right.
\end{eqnarray}
with the ratio $x=\frac{E_c}{2E_s}$. $E_c=a_c Z^2/A^{1/3}$ denotes the Coulomb energy and $E_s=a_s A^{2/3}(1-\kappa I^2)$ denotes the surface energy with isospin asymmetry $I= (N-Z)/A $. The coefficients $a_c=0.7092$ MeV, $a_s=17.4090$ MeV and $\kappa=1.5189$ are taken from WS4 \cite{WS4}. Together with the shell corrections $U_{\rm gs}$, $U_{\rm sad}$ and the macroscopic deformation energy $B_{\rm def}$ also from the WS4 predictions, the fission barrier heights for all bound heavy nuclei can be calculated by using
\begin{eqnarray}
B_{\rm f}^{\rm WS4}= B_{\rm f}^{\rm mac}-U_{\rm gs}+U_{\rm sad}-B_{\rm def},
\end{eqnarray}
neglecting the residual correction $\Delta B$. The influence of $\Delta B$ will be discussed later.

In this work, the saddle point of a nucleus is determined from the surface of the shell correction $U(\beta_2,\beta_4)$ based on the WS4 calculations in which the Strutinsky shell correction is obtained with the single-particle levels of an
axially deformed Woods-Saxon potential. As two examples, we show in Fig. 1 the contour plot of the shell correction surface for nuclei $^{208}$Pb and $^{238}$U. Here, other deformations such as $\beta_3$ and $\beta_6$ are neglected in the calculations.  From the contour plot, one can see that the shell correction $U_{\rm sad}$ at the saddle point is about 2.2 MeV for $^{208}$Pb and 1.0 MeV for $^{238}$U. In the calculations, we introduce a truncation for the macroscopic deformation energy, i.e. $B_{\rm def}\leqslant B_{\rm f}^{\rm mac}$, and we neglect the influence of $U_{\rm sad}$ for nuclei with $U_{\rm sad}<0$.

\begin{figure}
\includegraphics[angle=0,width=0.85\textwidth]{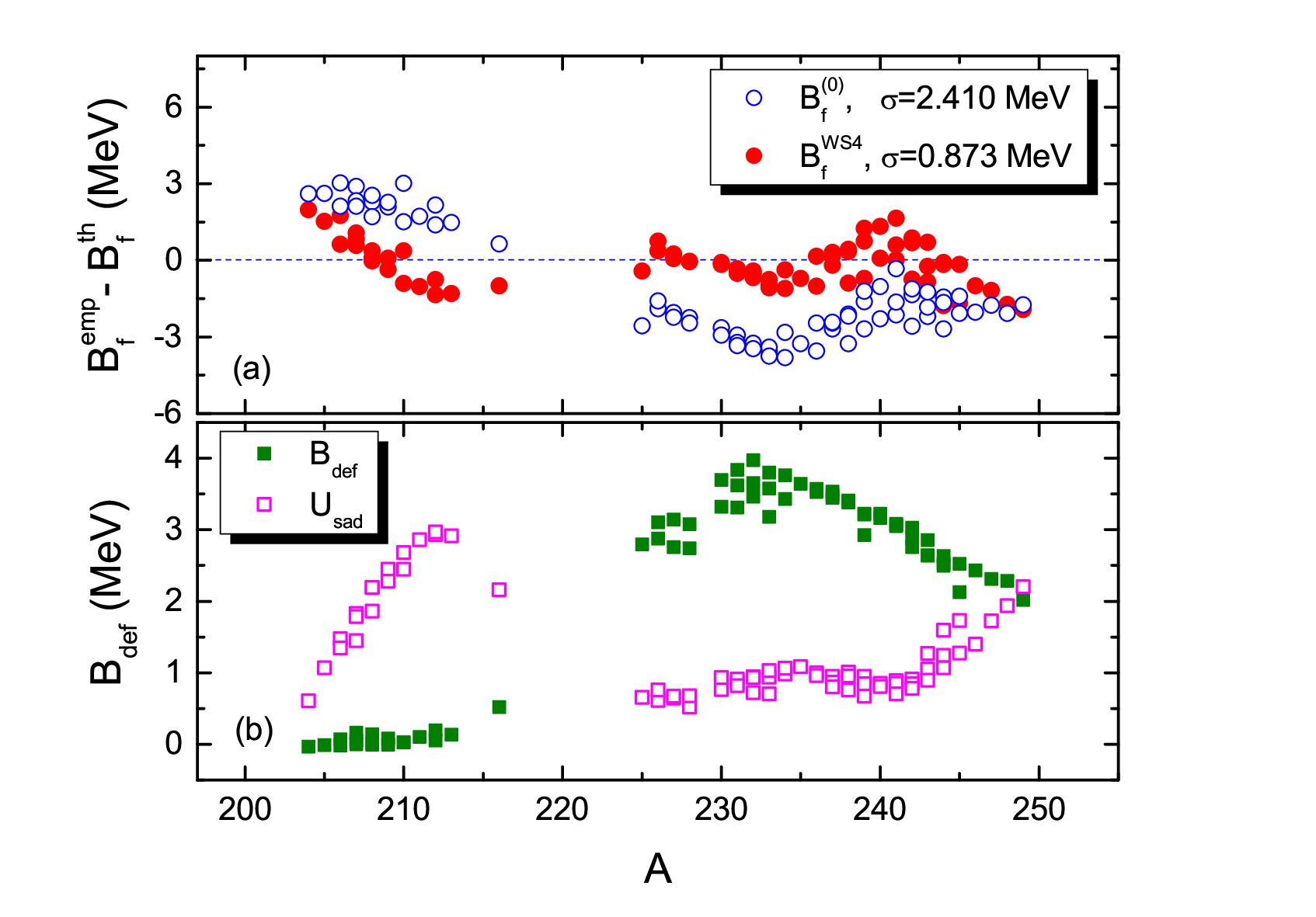}
\caption{(Color online) (a) Deviations between the empirical fission barrier heights $B_{\rm f}^{\rm emp}$ \cite{Cap09} and the model predictions. The open circles denote the results with $B_{\rm f}^{\rm (0)} = B_{\rm f}^{\rm mac}-U_{\rm gs}$ and the solid circles denote the results with Eq.(4). (b) Macroscopic deformation energies $B_{\rm def}$ (solid squares) and shell corrections at the saddle points $U_{\rm sad}$ (open squares) for these nuclei. }
\end{figure}

\begin{center}
\textbf{ III. RESULTS AND ANALYSIS }
\end{center}

\begin{figure}
\includegraphics[angle=0,width=0.75\textwidth]{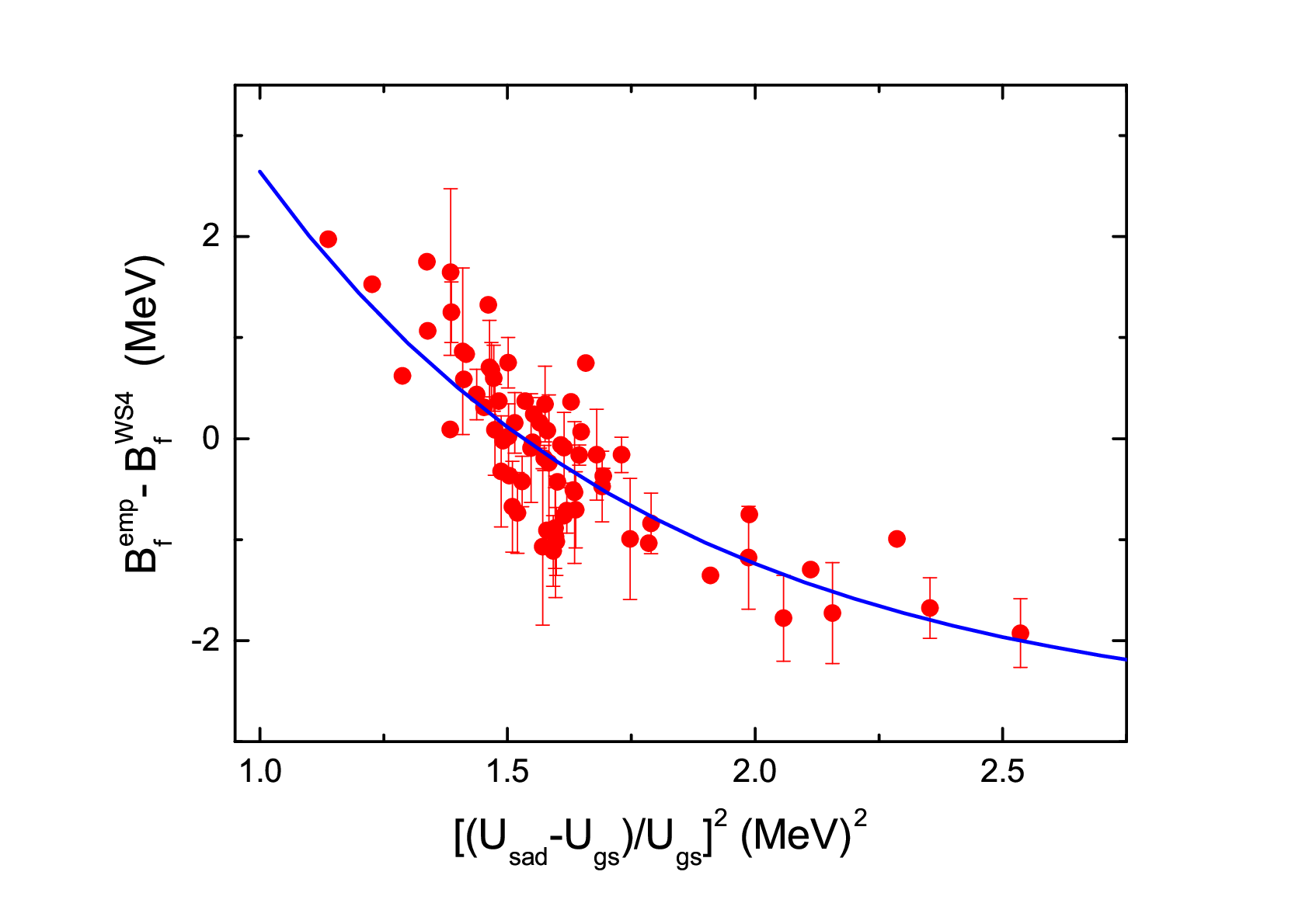}
\caption{(Color online) Deviation $B_{\rm f}^{\rm emp}-B_{\rm f}^{\rm WS4}$ as a function of the ratio square of the shell correction. The error bars denote the difference between the inner barrier height and the corresponding outer barrier height for actinides \cite{Cap09}. The solid curve denotes the results from Eq.(5).}
\end{figure}

\begin{figure}
\includegraphics[angle=0,width=0.75\textwidth]{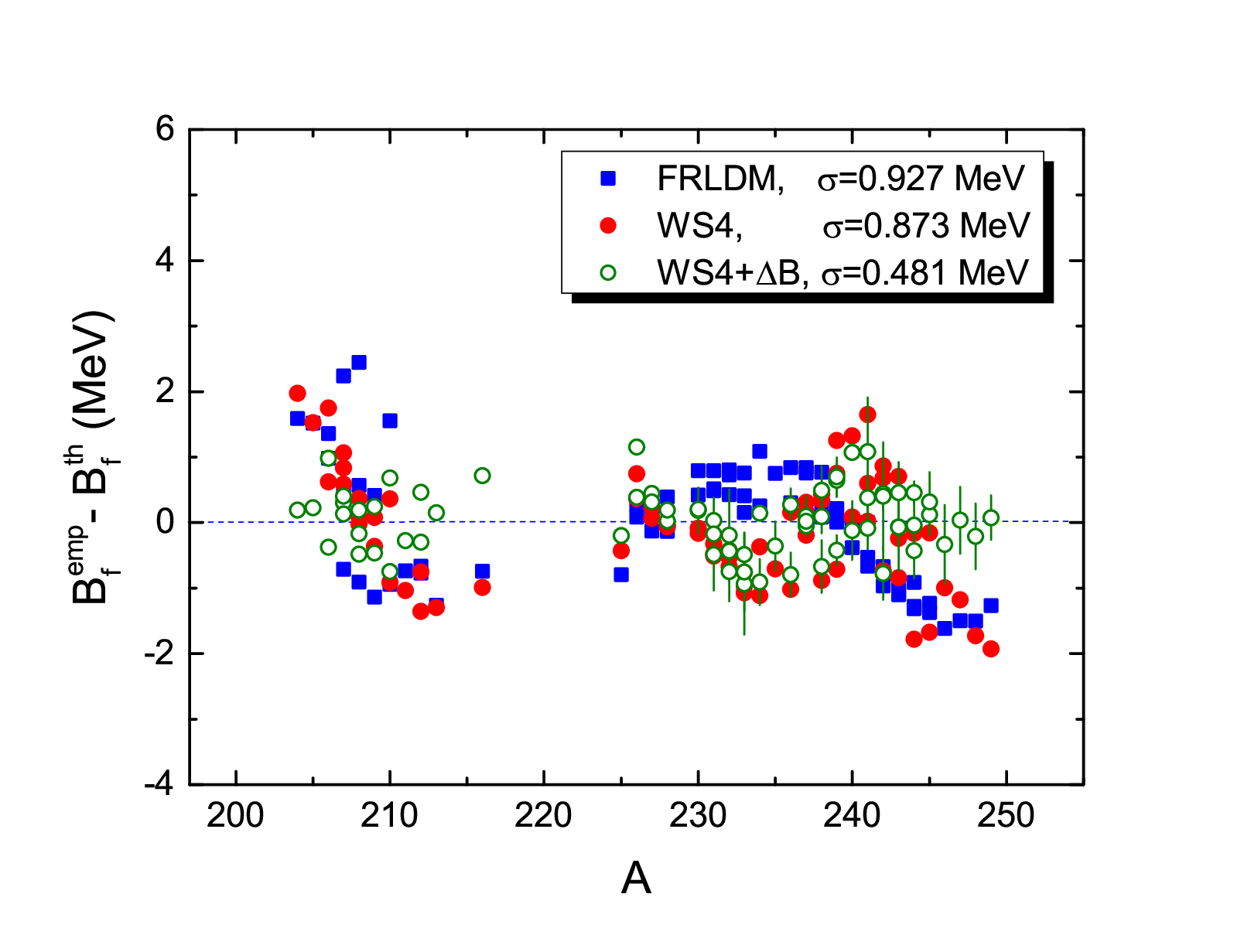}
\caption{(Color online) The same as Fig. 2(a), but with the results of the finite-range liquid-drop (FRLDM) model \cite{Moll15} and $B_{\rm f}^{\rm WS4}+\Delta B$ for comparison. The error bars denote the difference between the inner barrier height and the corresponding outer barrier height for actinides \cite{Cap09}}
\end{figure}

In Fig. 2(a), we show the deviations between the empirical fission barrier heights $B_{\rm f}^{\rm emp}$ \cite{Cap09} and the model predictions  for 71 nuclei with $Z\geqslant 82$. For actinides, we take the mean value of the inner and the outer barrier height as the value of $B_{\rm f}^{\rm emp}$ in the comparisons. One sees that with the saddle point shell correction $U_{\rm sad}$ and the macroscopic deformation energy $B_{\rm def}$ being taken into account, the root-mean-square (rms) deviation with respect to the fission barrier heights is significantly reduced from 2.410 MeV to 0.873 MeV. We note that for both of spherical nuclei and deformed nuclei, the fission barrier heights are generally better described by $B_{\rm f}^{\rm WS4}$. For doubly-magic nucleus $^{208}$Pb, the calculated $B_{\rm f}^{\rm (0)}=25.1$ MeV which is smaller than the empirical barrier height by 2.3 MeV. Considering $U_{\rm sad}=2.2$ MeV for $^{208}$Pb, we obtain $B_{\rm f}^{\rm WS4}=27.3$ MeV, which is very close to the empirical value. For deformed nucleus $^{238}$U, the calculated $B_{\rm f}^{\rm (0)}=9.2$ MeV which is higher than the empirical barrier by 2.9 MeV. With the macroscopic deformation energy of $B_{\rm def}=3.4$ MeV and $U_{\rm sad}=1.0$ MeV, one obtains $B_{\rm f}^{\rm WS4}=6.8$ MeV for $^{238}$U which is comparable with the empirical value. From Fig. 2(a), one sees that the values of $B_{\rm f}^{\rm emp}-B_{\rm f}^{(0)}$ are well divided into two parts: around 2 MeV for nuclei with $A \sim 210$ and about $-3$ MeV for $A>225$. To see the physics behind, the macroscopic deformation energies $B_{\rm def}$ and the saddle point shell corrections $U_{\rm sad}$ for these nuclei are shown in Fig. 2(b). One notes that for nuclei with $A\sim 210$, the values of $U_{\rm sad}$ are obviously larger than $B_{\rm def}$, whereas for $A>225$, $B_{\rm def} > U_{\rm sad}$ generally. The values of $U_{\rm sad}-B_{\rm def}$ are therefore well divided into two parts.

\begin{figure}
\includegraphics[angle=0,width=0.75\textwidth]{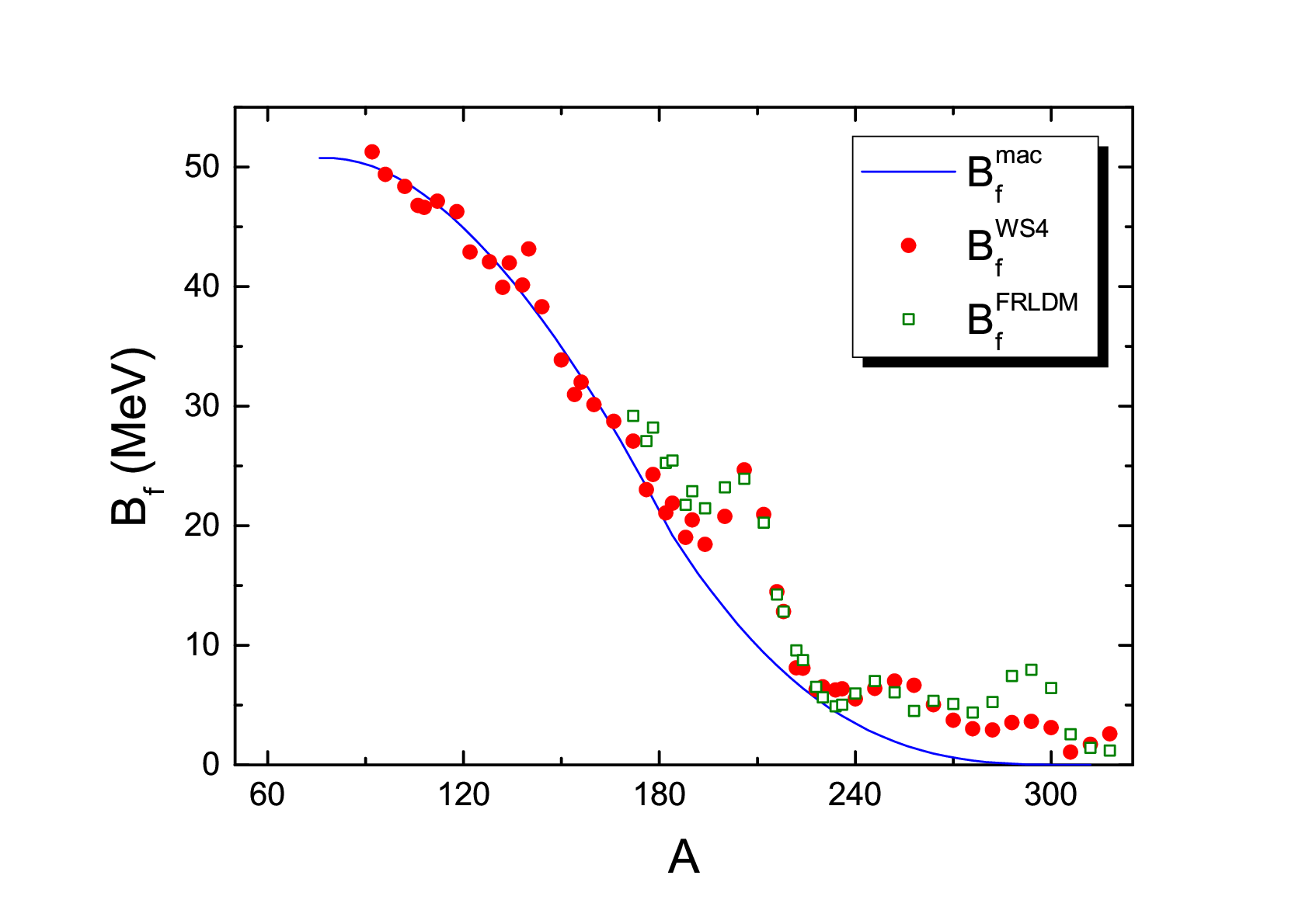}
\caption{(Color online) Fission barrier heights for even-even nuclei which lie on or are the closest
to the $\beta$-stability line (using Green's expression \cite{Green}) .}
\end{figure}

To further analyze the influence of the saddle point shell correction, we introduce an quantity $\delta U=(U_{\rm sad}-U_{\rm gs})/U_{\rm gs}$ to describe the relative value of the shell correction. In Fig. 3, we show the values of $B_{\rm f}^{\rm emp}-B_{\rm f}^{\rm WS4}$ as a function of $\delta U^2$. One sees that the difference between the empirical barrier heights and the model predictions systematically decrease with $\delta U^2$. Comparing Eq.(2) with Eq.(4), one notes that the residual correction $\Delta B$ is neglected in $B_{\rm f}^{\rm WS4}$. To better describe the fission barriers, we empirically write the residual correction as (in MeV),
\begin{eqnarray}
\Delta B \approx -2.8+ 19\exp(-\delta U^2/0.8).
\end{eqnarray}
In Fig. 4, we show the results of $B_{\rm f}^{\rm WS4}+\Delta B$ and those of FRLDM \cite{Moll15} for comparison. The rms deviation is further reduced to 0.481 MeV considering the residual correction given by Eq.(5).  We note that the trend of the results from FRLDM is similar to that from WS4 at the regions of $A\approx 210$ and $A>240$. In addition, we also note that the mean values of the fission barrier heights from the three macroscopic-microscopic approaches, i.e. $\langle B_{\rm f} \rangle =(B_{\rm f}^{\rm FRLDM}+B_{\rm f}^{\rm WS4}+B_{\rm f}^{\rm WS4+\Delta B })/3$, are also in good agreement with the empirical values, with an rms error of only 0.585 MeV. In the calculations of $ \langle B_{\rm f} \rangle$ we set a relatively larger weight for the WS4 model, considering its smaller rms error for describing known masses and the empirical $B_{\rm f}$. We would like to emphasize that the difference between the inner barrier heights and the outer ones for actinides ($Z\geqslant 90$) could result in some uncertainties in analyzing the model accuracy. Comparing with $B_{\rm f}^{\rm WS4}$, the rms deviation is reduced from 1.01 MeV to 0.77 MeV for the 45 inner barriers of actinides with $\Delta B$ being considered, and the corresponding value is reduced from 0.92 MeV to 0.48 MeV for the outer barriers.

With the proposed approach, we systematically study the fission barrier heights for stable nuclei. In Fig. 5, we show the predicted $B_{\rm f}$ for even-even nuclei which lie on or are the closest to the $\beta$-stability line (Green's expression \cite{Green}, $N-Z=0.4A^2/(A+200)$, is arbitrarily adopted). The solid curve denotes the results of the macroscopic fission barrier $B_{\rm f}^{\rm mac}$ given by Eq.(3). One sees that the macroscopic barrier height approaches zero for super-heavy nuclei. For nuclei around $^{208}$Pb, the predicted fission barrier heights with both FRLDM and WS4 models are significantly higher than those of $B_{\rm f}^{\rm mac}$ due to the strong shell effects. For actinides with $A\approx 240$, the results of WS4 are close to those of FRLDM, with values of about 6 MeV. For super-heavy nuclei around $^{294}$Cn, the results of FRLDM are higher than those of WS4 by about 4 MeV.

\begin{figure}
\includegraphics[angle=0,width=0.9\textwidth]{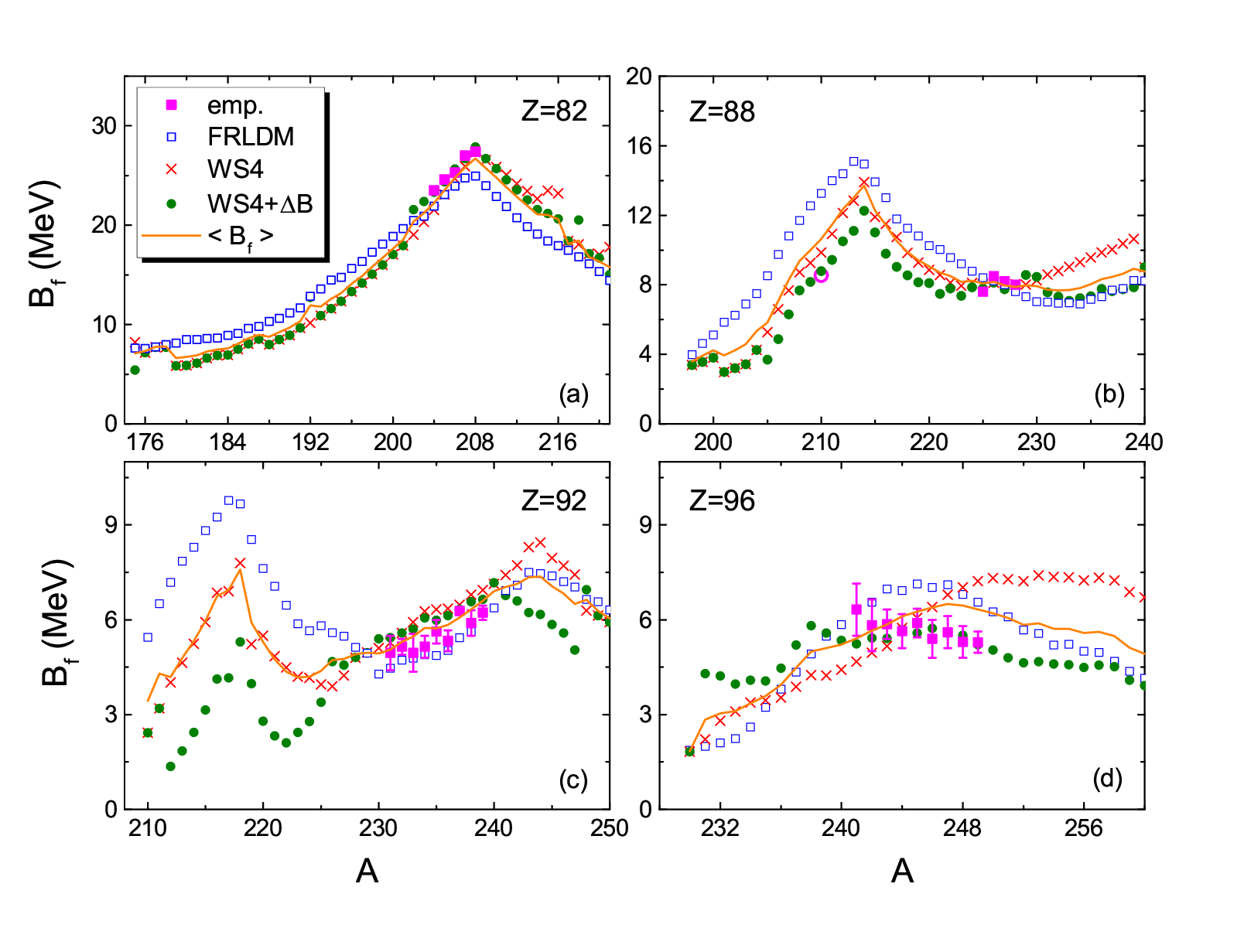}
\caption{(Color online) Predicted fission barrier heights for isotopic chains of Pb, Ra, U and Cm. The open squares denote the results of FRLDM. The crosses and the solid circles denote the results of WS4 with Eq.(4) and those together with the residual correction $\Delta B$, respectively. The solid curve denotes the mean value of the fission barrier heights $\langle B_{\rm f} \rangle$ from the three approaches. The open circle in (b) denotes the measured $B_{\rm f}$ for $^{210}$Ra taken from \cite{Ves24}. }
\end{figure}

In addition, with the proposed approach we simultaneously investigate the fission barrier heights of unstable nuclei. In Fig. 6, we show the predicted barrier heights for isotopic chains of Pb, Ra, U and Cm. The pink squares denote the empirical barrier heights taken from \cite{Cap09}. For Pb isotopes, the fission barriers evidently decrease with the increasing distance from the doubly-magic nucleus $^{208}$Pb. For Ra and U isotopes, all three approaches predict that there exists a peak at the neutron-deficient side with neutron number $N=126$. For $^{218}$U, the predicted barrier height $B_{\rm f}^{\rm WS4+\Delta B}=5.30$ MeV, while the result of FRLDM is 9.67 MeV. Very recently, the fission barrier heights for neutron-deficient nuclei $^{210}$Fr and $^{210}$Ra have been measured \cite{Ves24}, and the obtained data are $B_{\rm f}$($^{210}$Fr) = 10.67 MeV and $B_{\rm f}$($^{210}$Ra) = 8.54 MeV with uncertainty of 5\%. The predicted values of $B_{\rm f}^{\rm WS4+\Delta B}$ are 11.50 MeV and 8.78 MeV for $^{210}$Fr and $^{210}$Ra, respectively. From Fig. 6(b), one sees that the measured $B_{\rm f}$($^{210}$Ra) (open circle) can be well reproduced by using the WS4$+\Delta B$ calculations. For the neutron-rich side, the trend of fission barrier height is also strongly influenced by the shell corrections. In Fig. 7, we show the predicted $B_{\rm f}$ for nuclei with $Z=102$, 106, 119 and 120. For No and Sg isotopes, the peaks of the fission barrier heights at $N=152$ and $N=162$ can be evidently observed. The abrupt change of $\alpha$-decay energies at neutron number of 152 and 162 due to the shell effects can also be evidently observed for heavy and super-heavy nuclei \cite{Zhang12}.

For unknown super-heavy nuclei (SHN) with $Z=119$ and 120, the predicted fission barrier heights are presented in Fig. 7(c) and (d). One sees that for the SHN with $Z=119$ and $A=297$, the predicted barrier height with the FRLDM is 7.94 MeV, which is  higher than that of WS4 model by 2 MeV. We note that in the study of fusion reaction $^{48}$Ca+$^{238}$U \cite{Nisho12}, the negative of the shell correction energy (6.64 MeV) from the finite range droplet model (FRDM) \cite{Moll95} is taken as the fission barrier height of the compound nucleus, but multiplied with 0.7 in order to reproduce the maximum cross section for $^{283}$Cn(3n) measured at an excitation energy of 35.0 MeV. The predicted value of $B_{\rm f}^{\rm WS4}=4.20$ MeV for $^{286}$Cn, which is comparable with the result of FRDM multiplied with a factor of 0.7. If considering the reduction factor of 0.7 for FRLDM, the fission barrier heights from the two models are comparable for the SHN with $Z=119$ and $A=297$. For the SHN with $Z=120$, the largest value of $B_{\rm f}^{\rm WS4}=6.23$ MeV is located at $N=176$, which is consistent with the predictions of Warsaw's macroscopic-microscopic calculations \cite{Kow10}.

\begin{figure}
\includegraphics[angle=0,width=0.9\textwidth]{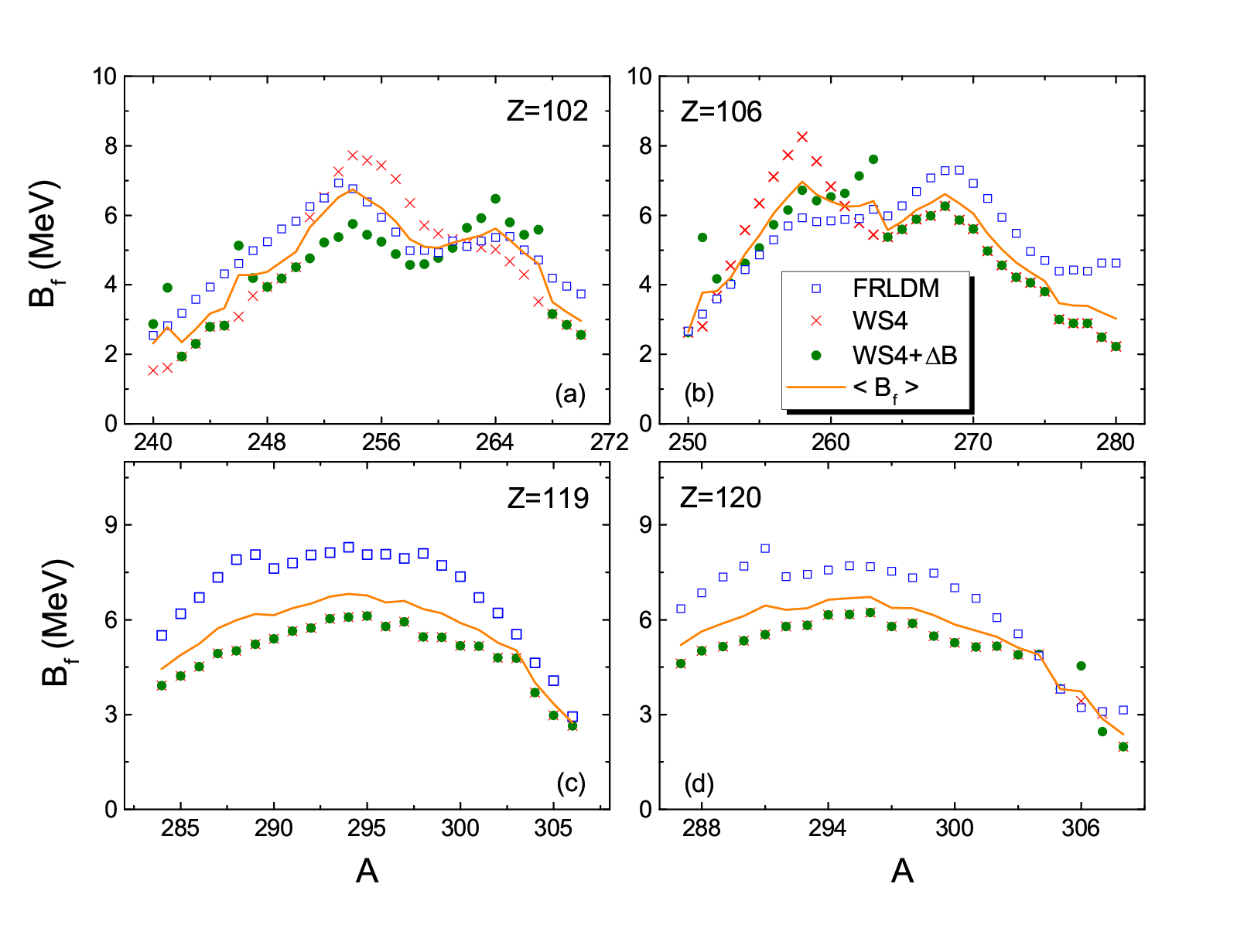}
\caption{(Color online) The same as Fig. 6, but for heavier nuclei.}
\end{figure}

\begin{center}
\textbf{IV. SUMMARY}
\end{center}

Based on the WS4 mass model with which the known masses can be reproduced with an rms error of $\sim0.3$ MeV and the known $\alpha$-decay energies of SHN can be reproduced with deviations smaller than 0.5 MeV, the fission barrier heights of heavy and superheavy nuclei are systematically studied. Considering the shell corrections, the macroscopic deformation energy and a phenomenological residual correction, the fission barrier heights for nuclei with $Z\geqslant 82$ can be well described, with an rms error of only 0.481 MeV. We note that in addition to the shell correction at the ground state, the shell correction at the saddle point and its relative value are also important for accurate description of the barrier height. From the predicted fission barriers for isotopic chains of Pb, Ra, No and Sg, we note that the influence of the shell effect on the barrier height is evident. For Ra and U isotopes, all three approaches predict that there exists a peak at the neutron-deficient side with $N=126$. For No and Sg isotopes, the peaks of the barrier heights at $N=152$ and $N=162$ can also be evidently observed. With the predicted fission barriers, the evaporation residual cross sections in the fusion reactions searching for new neutron-deficient isotopes \cite{Zhang24} and the reactions leading to the synthesis of super-heavy nuclei \cite{Ogan23} could be analyzed more accurately.

\begin{center}
\textbf{ACKNOWLEDGEMENTS}
\end{center}
This work was supported by National Natural Science Foundation of China (Nos. 12265006, U1867212), Guangxi Natural Science Foundation (No. 2017GXNSFGA198001). The table of the fission barriers with the WS4 model is available from  http://www.imqmd.com/mass/BfWS4.txt

\end{document}